\documentclass[twocolumn,tighten]{aastex62}

\usepackage{amsmath}
\usepackage{multirow}
\usepackage{booktabs}

\begin{document}

\title{Imaging the Dusty Substructures due to
Terrestrial Planets in Planet-forming Disks with ALMA and the Next
Generation Very Large Array}

\author{Sarah K. Harter}
\affiliation{Department of Physics and Astronomy, California State University Northridge, 18111 Nordhoff Street, Northridge, CA 91330, USA}

\author{Luca Ricci}
\affiliation{Department of Physics and Astronomy, California State University Northridge, 18111 Nordhoff Street, Northridge, CA 91330, USA}

\author{Shangjia Zhang}
\affiliation{Department of Physics and Astronomy, University of Nevada Las Vegas, Las Vagas, NV 89154, USA}

\author{Zhaohuan Zhu}
\affiliation{Department of Physics and Astronomy, University of Nevada Las Vegas, Las Vagas, NV 89154, USA}

\correspondingauthor{Luca Ricci}
\email{luca.ricci@csun.edu}

\begin{abstract}
\noindent We present simulations of the capabilities of the Atacama Large Millimeter/submillimeter Array (ALMA) and of a Next Generation Very
Large Array (ngVLA) to detect and resolve substructures due to terrestrial planets and Super-Earths in nearby planet-forming disks. 
We adopt the results of global 2-D hydrodynamical planet-disk simulations that account for the dynamics of gas and dust in a disk with an embedded planet. Our simulations follow the combined evolution of gas and dust for several thousand planetary orbits. We show that long integrations (several tens of hours) with the ngVLA can detect and spatially resolve dust structures due to low-mass rocky planets in the terrestrial planet formation regions of nearby disks (stellocentric radii $r = 1 - 3$ au), under the assumption that the disk viscosity in those regions is low ($\alpha \le 10^{-5}$). ALMA is instead unable to resolve these structures in these disk regions. We also show that high-resolution ngVLA observations separated by several days to few weeks would allow to detect the proper motion of the azimuthally asymmetric structures expected in the disk regions of terrestrial planet formation. 
\end{abstract}

\section{Introduction}

The mutual interactions between planets and the parental disk play a prominent role in setting the main physical properties of planetary systems. Planets are made of solids and gas which come from the disk, and the mutual gravitational interactions between a newly born planet and the material in the disk can strongly affect the history of its location within the planetary system \citep[for reviews see][]{Kley:2012,Baruteau:2014}. 

Recent high angular resolution observations at both optical/near-infrared and sub-millimeter/millimeter wavelengths have shown that \textit{substructures} in disks which resemble the predictions from planet-disk interactions theory are ubiquitous in young nearby disks \citep[e.g.,][]{Juhasz:2015,Dipierro:2018,Andrews:2018}. 
The detailed modeling of these substructures, especially rings and gaps, is often carried on via hydrodynamical simulations to infer key properties of the putative young planets, such as mass and orbital radius.
These studies show that the rings and gaps seen by ALMA from observations of the disk dust continuum likely trace a population of relatively massive young planets, with masses higher than $10~M_{\oplus}$, and relatively far from the host stars, at orbital radii greater than $10$ au \citep{Zhang:2018,Long:2018,Lodato:2019}.

Owing to different observational biases, the populations of these young planets and of the detected mature exoplanets have a very scarce overlap in terms of the main planetary and orbital properties, such as for example planet mass and orbital radius \citep[see Figure~21 in][]{Zhang:2018}. This strongly hinders comparisons between the properties of these young vs mature exoplanet populations which are key to unveil the processes of formation and early evolution of planetary systems \citep[see discussion in][]{Lodato:2019}. Future improvements on the observational capabilities regarding both direct and indirect methods of exoplanet detection will be necessary to inform theories of planet formation and evolution. 

Over the past several years, the astronomy community has shown strong interest for a future large area mm/radio interferometric array optimized for imaging at milliarcsecond (mas) resolution, which would be ideal for observations of protoplanetary disks, among other sources in the sky. This \textit{Next Generation Very Large Array} \citep[ngVLA,][]{Murphy:2018} would operate at frequencies of 1.2-116 GHz with approximately 10 $\times$  the collecting area of the current \textit{Very Large Array} (VLA) and \textit{Atacama Large Millimeter/submillimeter Array} (ALMA) interferometers, and with more than 10 $\times$ longer baselines that would yield mas-resolution at the highest frequencies. 

With the expected angular resolution and sensitivity at mm/cm wavelengths, the ngVLA would be able to spatially resolve structures in the inner terrestrial planet formation regions (1 au at the distance of the closest star forming regions, i.e. 140 pc, corresponds to an angular scale of about 7 mas). 
In this study we aim to quantify the capabilities of the ngVLA to detect and spatially resolve disk structures due to the planets with masses lower than $10~M_{\oplus}$ at orbital separations between 1 and 3 au from the host star, which are more in line with the properties of the bulk of the population of mature exoplanets detected so far. 

\citet{Ricci:2018} already tested the capabilities of the ngVLA to image disk substructures due to planets in the dust continuum emission of nearby disks.   However, the simulations that were run for that study followed the evolution of their disk-planet systems for less than 2000 planetary orbits. The models presented here follow the disk evolution for a larger number of orbits, giving more time for low-mass terrestrial planets to open gaps in the dust density of low-viscosity disks, which can be detected by the ngVLA \citep[see, e.g.][]{Dong:2018}.

Section~\ref{sec:methods} presents the methods used for this study, describing the disk physical models used to simulate the disk-planet interactions, the procedure used to obtain the model synthetic images at the ngVLA and ALMA wavelengths, and the method used to simulate the results of future ngVLA and ALMA observations. Section~\ref{sec:results} outlines the results of this investigation, Section~\ref{sec:discussion} presents a discussion of the main results, whereas Section~\ref{sec:conclusions} summarizes the conclusions of this work.

\section{Methods: from disk simulations to ngVLA and ALMA observations}
\label{sec:methods}

We describe here the methods used to simulate observations with the ngVLA and ALMA for the dust continuum emission of disks with embedded low-mass terrestrial planets.

\subsection{Disk model}
\label{sec:model}

The models used for this investigation were obtained via 2D hydrodynamical planet–disk simulations
using the modified version of the grid-based code FARGO
\citep{Masset:2000} called Dusty FARGO-ADSG \citep{Baruteau:2008a, Baruteau:2008b,Baruteau:2016}.
This code accounts for the dynamics of gas and dust in a disk with one or more embedded planets. In particular, it can calculate the gravitational interaction between gas and dust with the planet, as well as the aerodynamical gas-drag on the dust.

We refer to \citet{Zhang:2018} for a detailed description of the disk models and the method to extract synthetic images for the dust continuum at different wavelengths for a given model.
In this work, we initialized the disk structure assuming azimuthal symmetry with a  power-law radial dependence for the gas surface density:
\begin{equation}
    \Sigma {_g}(r) = \Sigma _{g,0}(r/r{_0}) ^{-1},
\end{equation}
where $r$ is the stellocentric radius in the disk, and $r_0$ is an arbitrary radius, that we assume to correspond to the orbital radius $r_p$ of the planet in the simulations.  
The normalization factor for the gas surface density $\Sigma_{g,0}$ is an input parameter for our models, and the values considered for the initial conditions of the hydrodynamical simulations presented here are $\Sigma_{g,0} = 100, 300, 600, 1200, 2400$ g/cm$^2$.
Observations of young disks conducted so far do not provide firm constraints on the gas surface density at stellocentric radii $r < 5 - 10$ au. The relatively large range of $\Sigma_{g,0}$ values sampled in this work reflects the large spread estimated for dust disk masses and disk radii in nearby star forming regions \citep[e.g.,][]{Andrews:2013,Andrews:2018a}. For reference, our largest value of $\Sigma_{g,0}$ is a factor of about 1.4 greater than the gas surface density at 1 au in the Minimum Mass Solar Nebula (MMSN) disk model presented in \citet{Hayashi:1981}, and a factor of 4 lower than in the MMSN model in \citet{Desch:2007}.

These models also assume the locally isothermal equation of state, where the temperature at stellocentric radius $r$ is given by $T(r) = T{_0}(r/r{_0})^{-1/2}$.  The disk temperature $T$ is related to the disk pressure scale height $h$ via $h/r = c_{s}/v_{\phi}$, where $c_{s} = \sqrt{RT/\mu}$ is the local sound speed of the gas ($\mu = 2.35$ is the adopted value for the mean molecular weight), and $v_{\phi}$ is the local rotation velocity of the gas, assumed to be keplerian.  
The disk aspect ratio at the orbital radius $r_{p} = r_0$ of the planet is assumed to be $h/r~(r_{p}) = 0.03$, and the models consider a Shakura-Sunyaev viscosity with an $\alpha$ parameter of $10^{-5}$ constant across the disk \citep{Shakura:1973}. 

For the luminosity of the central star we considered two values of $L_{\star} = 1 $ and $10~L_{\odot}$. In these models, the stellar luminosity $~L_{\star}$ determines the temperature across the disk via

\begin{equation}
    T_{d}(r) = \left(\dfrac{\phi L_{\star}}{8 \pi r^{2} \sigma}\right)^{1/4},
\label{eq:temp}
\end{equation}

where $\sigma$ is the Stefan-Boltzmann constant, and $\phi$ is a factor, taken to be 0.02 from \citet{DAlessio:2001}, which quantifies the effect of the flaring of the disk on its heating from the stellar light. 




We assume that the initial dust surface density of the disk is $1/100$ of the gas surface density, and the dust is characterized by a grain size distribution $n(s) \propto s^{-3.5}$, $s$ being the radius of the grain, with maximum grain size of 1 cm. Since our dust component uses open boundary conditions, the dust-to-gas mass ratio for the whole disk can change with time. Dust and gas experience both gravitational forces and the dust feels aerodynamic drag forces induced by gas. The Stokes number for particles, which controls the level of aerodynamical coupling, is given by $\rm{St} = t_{\rm{stop}}\Omega = \pi s \rho_{p}/2\Sigma_{g}$, where $t_{\rm{stop}}$ is the stopping time, $\Omega$ the angular velocity of the gas, and $\rho _{p}$ the solid density of dust particles. The combined temporal evolution of dust and gas in the hydrodynamical code adopted in this work is described in more detail in \citet{Zhang:2018}.

The numerical grid extends from $0.1 r_{0}$ to $10 r_{0}$ in the radial direction and from 0 to $2\pi$ in the azimuthal direction, and is sampled by 750 and 1024 grid points in the radial and azimuthal directions, respectively. 

In these models, the parameter $q$ represents the ratio between the planet mass $M_p$ and the stellar mass $M_{\star}$ \citep[see][]{Zhang:2018}. We have investigated two families of models, one with $q = M_{p}/M_{\star} = 1~M_{\oplus}/M_{\odot}$ and with $10~M_{\oplus}/M_{\odot}$. The first family corresponds to having an Earth-mass planet orbiting a Solar-mass star, or a Mars-mass planet around a M-type star with mass of $0.1~M_{\odot}$. The second family corresponds to having a $10~M_{\oplus}$-mass Super Earth around a Solar-mass star, or an Earth-mass planet around a star with mass of $0.1~M_{\odot}$. 

The simulations with $q = 1~M_{\oplus}/M_{\odot}$ were run for up to 7000 planet orbits with the planet located at either $r_p =$ 2 or 3 au from the host star. For the simulations with $q = 10~M_{\oplus}/M_{\odot}$, the simulations were run for up to 5000 planet orbits with the planet at $r_p =$ 1 or 3 au. A summary of the values for the model parameters investigated in this study is presented in Table~\ref{table:1}.

\begin{figure*}[ht!]
\begin{center}
\includegraphics[scale=0.55]{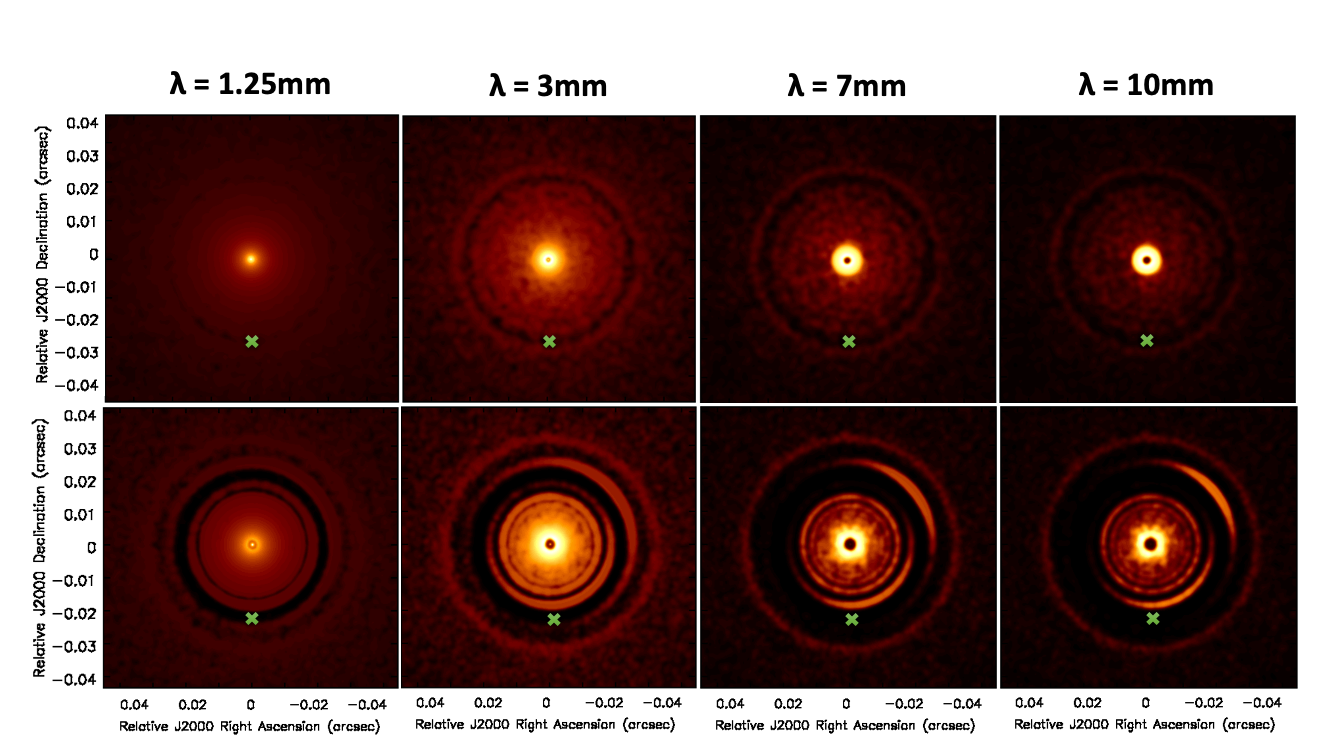}
\end{center}
\caption{\footnotesize {Model synthetic images for the dust continuum emission at 1.25, 3, 7 and 10 mm, respectively, for two of the systems considered in this study. Top row) Disk model with a planet at 3 au from the host star and a planet-to-star mass ratio of $1~M_{\oplus}/M_{\odot}$ at 7000 planet orbits as discussed in Section~\ref{sec:results3.2}. The planet location in each panel is marked with a green cross. The integrated flux densities for this model are 42.0, 4.2, 0.32 and 0.099 mJy at the wavelengths of 1.25, 3, 7 and 10 mm, respectively. Bottom row) Disk model with a planet at 3 au from the host star and a planet-star mass ratio of $10~M_{\oplus}/M_{\odot}$ at 5000 planet orbits as discussed in Section~\ref{sec:results3.1}. The integrated flux densities for this model are 18.7, 2.05, 0.170 and 0.055 mJy at the wavelengths of 1.25, 3, 7 and 10 mm, respectively.  
}}
\label{fig:models}
\end{figure*}

The synthetic maps for the dust continuum emission of the disk models were derived using the methods described in \citet{Zhang:2018}, and were produced at wavelengths of 1.25, 3, 7 mm and 1 cm, and two representative cases are shown in Figure~\ref{fig:models}.

\begin{table*}
\centering
\begin{tabular}{crccccc}  
\hline
\hline
 $q$ [$M_{\oplus}/M_{\odot}$]  & \multicolumn{6}{c}{Planet orbital radius [au]} \\ \cline{2-6}
 & \multicolumn{5}{c}{$L_{\star} = 10~L_{\odot}$} & $L_{\star} = 1~L_{\odot}$ \\
 \cline{2-6}
 & $\Sigma _{g,0}^{\rm{init}}[\rm{g/cm}^{2}] = 100$ & 300 & 600 & 1200 & 2400 & $\Sigma _{g,0}^{\rm{init}} = 300~\rm{g/cm}^{2}$\\
 \hline
 1 & .. & 2, 3 & .. & .. & .. & 3\\
 10 & 3 & 1, 3 & 3 & 3 & 3 & 1\\
 \hline
 \hline
\end{tabular}
\caption{\footnotesize{Values of the model parameters investigated in this study.}}
\label{table:1}
\end{table*}

\subsection{Simulations of the ngVLA and ALMA observations}

The synthetic maps obtained according to the methods outlined in Section~\ref{sec:model} were converted into predictions for future observations with the ngVLA (at $\lambda = 3, 7, 10$ mm) and ALMA ($\lambda = 1.25$ mm) using the CASA software package\footnote{\texttt{https://casa.nrao.edu/}}.
We adopted the same procedure as in \citet{Ricci:2018}, in which the ngVLA simulations are performed using the \texttt{SIMOBSERVE} task to generate the visibility dataset in the $(u,v)$ Fourier space, and the \texttt{SIMNOISE} task to add the noise by corrupting the visibilities. According to the specifics of the reference design described in \citet{Selina:2018}, the rms noise levels on the ngVLA maps presented in this paper correspond to on-source integrations times of $\approx 70-80$ hours.

For the ngVLA simulations we considered the original ngVLA Rev B array configuration
distributed across the US Southwest and Mexico. This array configuration includes 214 antennas of 18-meter diameter, to baselines up
to 1000\,km \citep{Selina:2018}. In our simulations we did not include the 30 additional antennas distributed to baselines up to 9000\,km, as the signal from the sources considered here would be very faint for these very long baselines. For the array configuration of the ALMA observations, we used the \texttt{alma.out28.cfg} antenna position file available in the
CASA package, which incorporates the longest 16 km baselines in the ALMA array.

For the imaging of the visibilities we employed the \texttt{CLEAN} algorithm with Briggs weighting, and adjusted the robust parameter to give a reasonable synthesized beam and noise performance.
In particular, the ALMA images were computed
with a Briggs weighting with robust parameter $R = -2$ (uniform weighting), while for the ngVLA we chose
$R = -1$. For the deconvolution we employed a multiscale clean approach with scales of 0, 7 and 30 pixels to better recover compact emission at both high brightness and large diffuse
structures in the disk models.
The disk center was located at a declination of $+35.0$ degrees for the ngVLA simulations and $-24.0$ degrees for the ALMA simulations, and the assumed distance is 140~pc, similar to the distance of nearby star forming regions such as Taurus or Ophiuchus.


\section{Results}
\label{sec:results}

To facilitate the discussion of the results of the simulations with different parameters for the disk, planet and star, we start by first presenting the results of one specific model in Section~\ref{sec:results3.1}. In the following sections, we discuss separately the results when changing the planet-to-star mass ratio, disk surface density, planet orbital radius, and stellar luminosity, respectively.

\subsection{Reference model}
\label{sec:results3.1}

\begin{figure*}[ht!]
\begin{center}
\includegraphics[scale=.65]{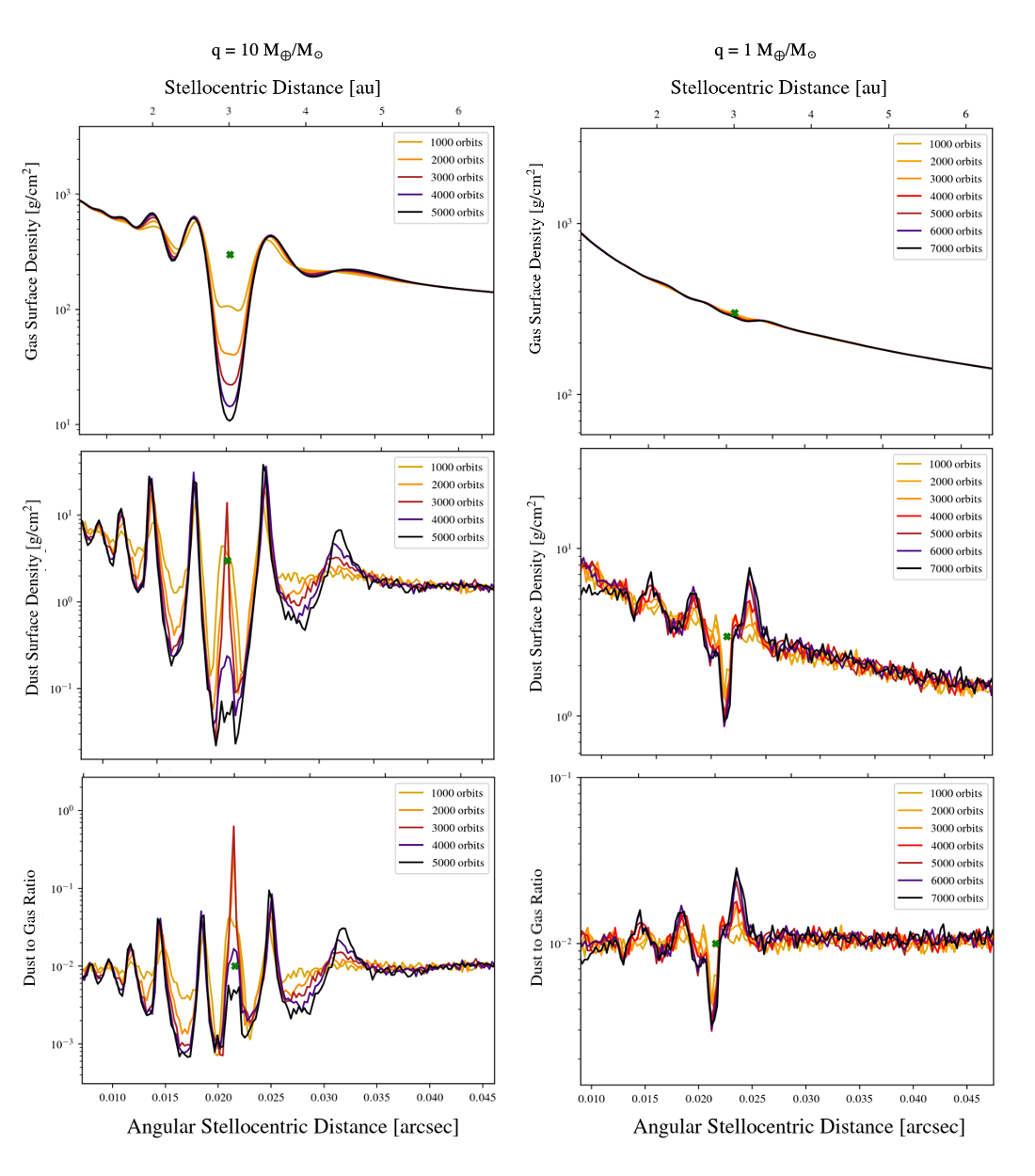}
\end{center}
\caption{\footnotesize {The azimuthally averaged radial profiles  of the gas (top row) and dust density (middle) and dust-to-gas mass ratio (bottom) for models with a planet-to-star mass ratio $q = 10~M_{\oplus}/M_{\odot}$ (left) and $1~M_{\oplus}/M_{\odot}$ (right) with a planet located at 3 au, and an initial gas surface density of 300 g/cm$^2$ at the location of the planet. Different radial profiles for different snapshots during the simulations are shown in different colors as labeled on each panel. In each panel, the abscissa of the green cross represents the stellocentric distance of the planet, while its ordinate corresponds to the value on the y-axis at the beginning of the simulations.}
}
\label{fig:gasdustdens}
\end{figure*}

\begin{figure*}
\begin{center}
\includegraphics[scale=.45]{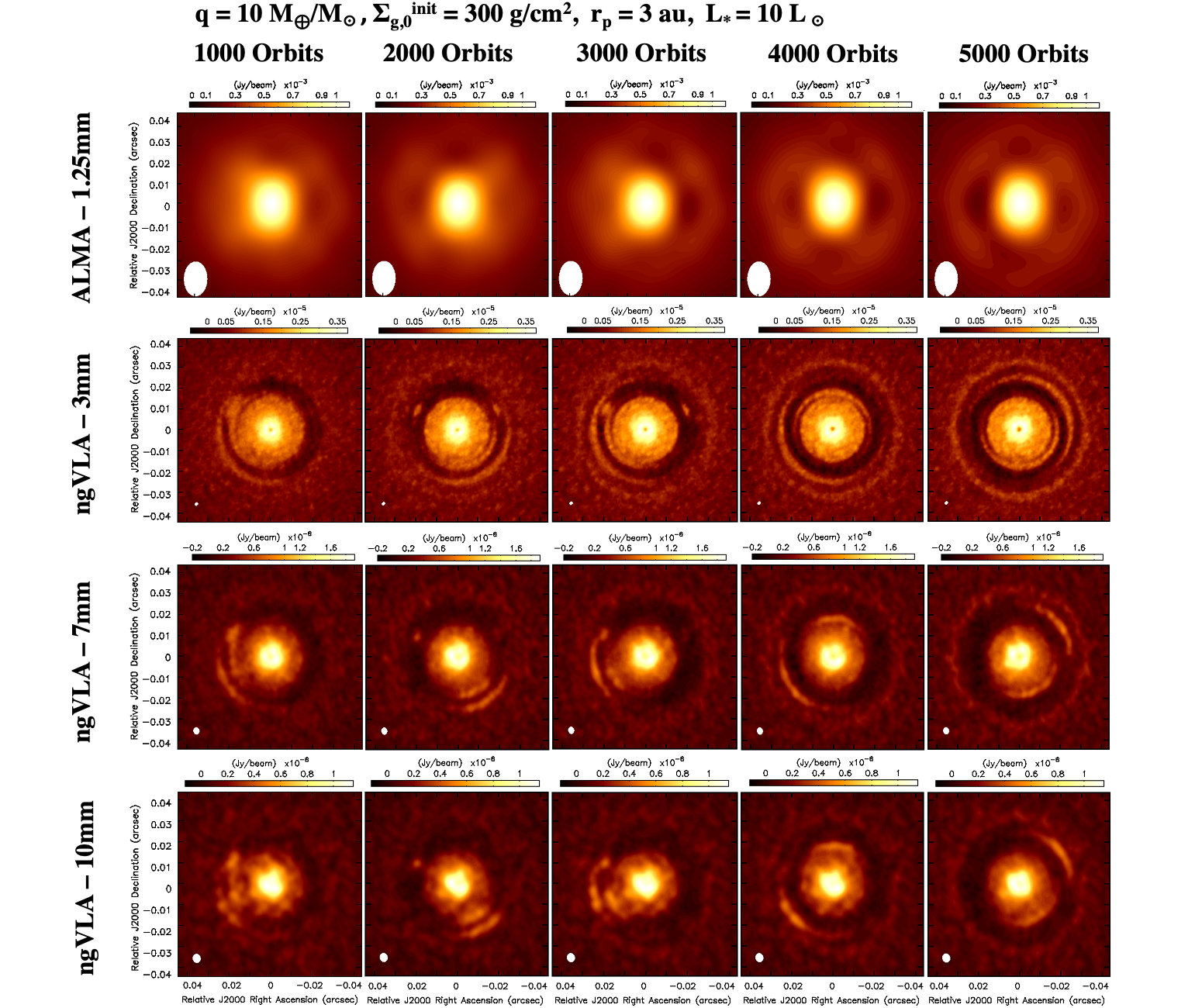}
\end{center}
\caption{\footnotesize {ALMA and ngVLA simulated observations for the dust continuum emission at 1.25 mm (ALMA), 3 mm, 7 mm, and 10 mm (ngVLA) for the disk model with a planet at 3 au from the host star and planet-to-star mass ratio of $10~M_{\oplus}/M_{\odot}$, stellar luminosity $L_{\star} = 10~L_{\odot}$, and disk with initial surface gas density of 300 g/cm$^2$ at the location of the planet. This is the reference model introduced in Section~\ref{sec:results3.1}. Each column represents the specific snapshot of the simulation in terms of number of planet orbits, from left to right of 1000, 2000, 3000, 4000, and 5000 orbits, respectively. 
Each row represents simulations of either ALMA or ngVLA observations at the wavelengths specified in the figure.
The resulting FWHM size of the synthesized beams are 18\,mas (2.5 au) $\times$ 12\, mas (1.7 au) at a wavelength of 1.25 mm,  1.7\,mas (0.23 au) $\times$ 1.3\, mas (0.19 au) at 3 mm, 3.4\,mas (0.47 au) $\times$ 2.7\, mas (0.38 au) at 7 mm, and 4.2\,mas (0.59 au) $\times$ 3.7\, mas (0.52 au) at 10 mm, respectively. The resulting rms noise on the maps are 2.8 $\mu$Jy\,beam$^{-1}$ at 1.25 mm, 86 nJy\,beam$^{-1}$ at 3 mm, 35 nJy\,beam$^{-1}$ at 7 mm, and 23 nJy\,beam$^{-1}$ at 10 mm. The disk model has integrated fluxes of 28.3 mJy at 1.25 mm, 6.6 mJy at 3 mm, 0.53 mJy at 7 mm, and 0.16 mJy at 10 mm, respectively, calculated at the 5000 orbits snapshot.}}
\label{fig:superearth_lstar10_rgap3_300g}
\end{figure*}

\begin{figure*}
\begin{center}
\includegraphics[scale=.45]{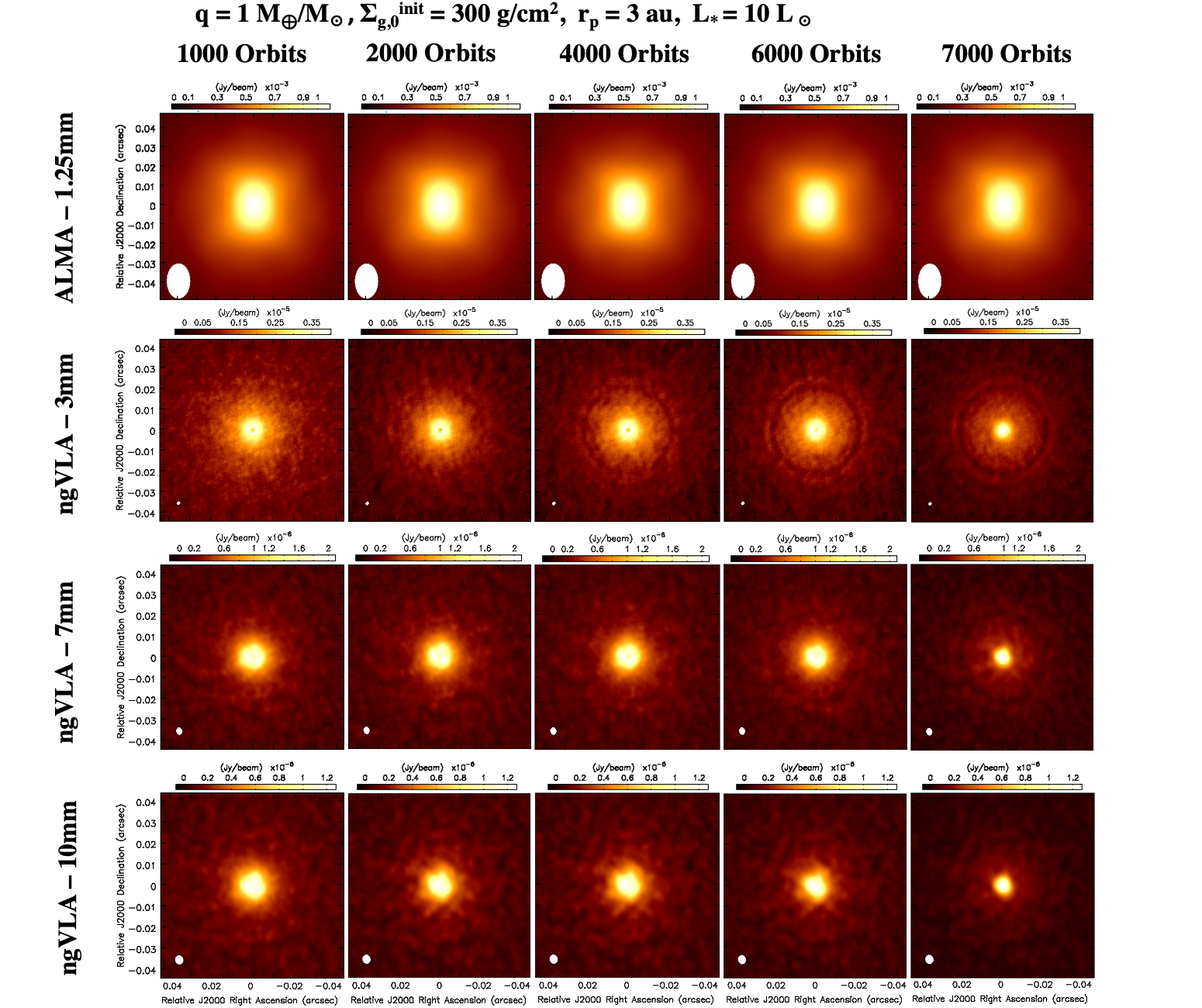}
\end{center}
\caption{\footnotesize {ALMA and ngVLA simulated observations for the dust continuum emission at 1.25 mm (ALMA), 3 mm, 7 mm, and 10 mm (ngVLA) for the disk models with a planet with planet-to-star mass ratio of $1~M_{\oplus}/M_{\odot}$ at 3 au from the host star, stellar luminosity $L_{\star} = 10~L_{\odot}$, and initial gas surface density of 300 g/cm$^2$ at the location of the planet. This is the model introduced in Section~\ref{sec:results3.2}. Each column represents the specific snapshot of the simulation in terms of planet orbits, from left to right, of 1000, 2000, 4000, 6000, and 7000 orbits. The resulting synthesized beams have FWHM $=$ 18\,mas (2.5 au) $\times$ 12\, mas (1.7 au) at 1.25 mm,  1.7\,mas (0.23 au) $\times$ 1.3\, mas (0.19 au) at 3 mm, 3.4\,mas (0.47 au) $\times$ 2.7\, mas (0.38 au) at 7 mm, and 4.2\,mas (0.59 au) $\times$ 3.7\, mas (0.52 au) at 10 mm. The resulting rms noise on the maps are 2800 nJy\,beam$^{-1}$ at 1.25 mm, 86 nJy\,beam$^{-1}$ at 3 mm, 35 nJy\,beam$^{-1}$ at 7 mm, and 23 nJy\,beam$^{-1}$ at 10 mm. The disk model has integrated fluxes of 29.0 mJy at 1.25 mm, 4.70 mJy at 3 mm, 0.523 mJy at 7 mm, and 0.157 mJy at 10 mm, respectively, calculated at the 7000 orbits snapshot.
}}
\label{fig:earth_lstar10_rgap3_300g}
\end{figure*}

\begin{figure*}
\begin{center}
\includegraphics[scale=.6]{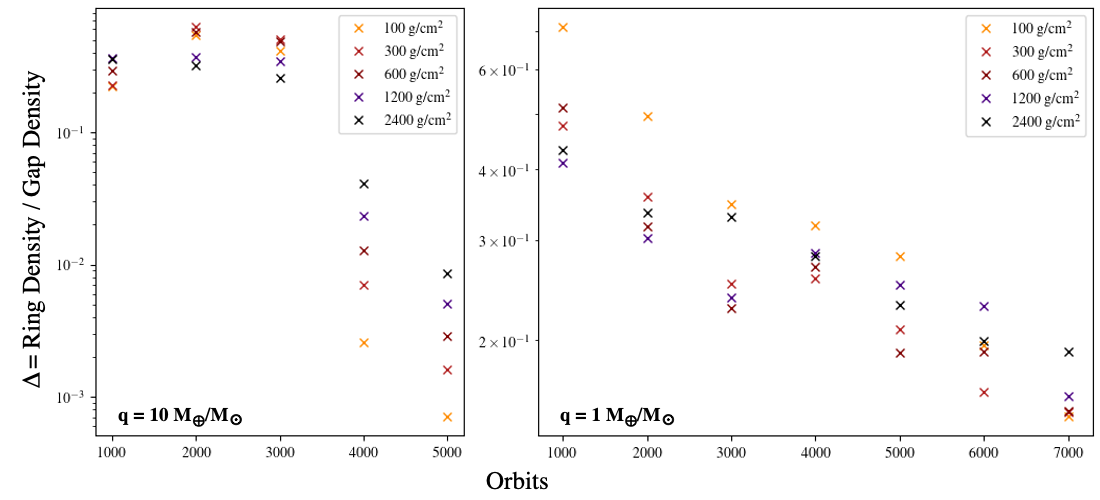}
\end{center}
\caption{\footnotesize {Temporal evolution for the $\Delta$ parameter, defined as the ratio between the average dust density in the gap at the orbital radius of the planet and the average density of the peaks of the two closest rings from the planet. Models are with $q = 10~M_{\oplus}/M_{\odot}$ (left panel) and $1~M_{\oplus}/M_{\odot}$ (right) with planet at an orbital radius of 3 au, and for several initial gas surface densities at the stellocentric radius of the planet, as labeled in each panel.}}
\label{fig:depletion}
\end{figure*}

\begin{figure*}
\begin{center}
\includegraphics[scale=.45]{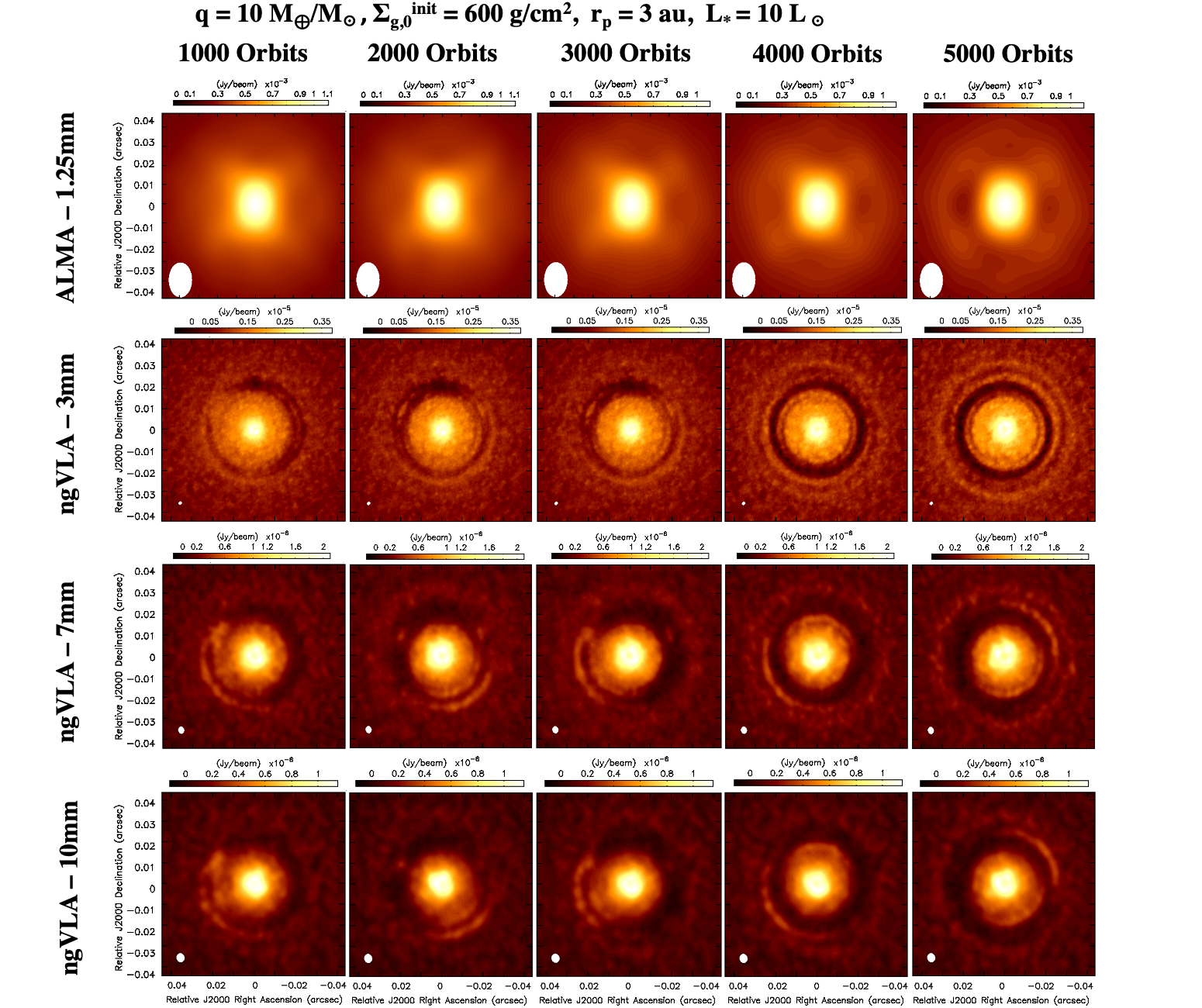}
\end{center}
\caption{\footnotesize {ALMA and ngVLA simulated observations for the dust continuum emission at 1.25 mm (ALMA), 3 mm, 7 mm, and 10 mm (ngVLA) for the disk model with a planet with planet-to-star mass ratio of $10~M_{\oplus}/M_{\odot}$ at 3 au from the host star, stellar luminosity $L_{\star} = 10~L_{\odot}$, and an initial gas surface gas density of 600 g/cm$^2$ at the location of the planet. This model is discussed in Section~\ref{sec:results3.3}. Each column represents the specific orbit from left to right of 1000, 2000, 3000, 4000, and 5000 orbits. The resulting synthesized beam has a FWHM $=$ 18\,mas (2.5 au) $\times$ 12\, mas (1.7 au) for 1.25 mm,  1.7\,mas (0.23 au) $\times$ 1.3\, mas (0.19 au) for 3 mm wavelength, 3.4\,mas (0.47 au) $\times$ 2.7\, mas (0.38 au) for 7 mm wavelength, and 4.2\,mas (0.59 au) $\times$ 3.7\, mas (0.52 au) for 10 mm wavelength  at the assumed distance of 140pc for our disk model. The resulting rms noise on the ngVLA map is 2800 nJy\,beam$^{-1}$ for 1.25 mm wavelength, 86 nJy\,beam$^{-1}$ at 3 mm wavelength, 35 nJy\,beam$^{-1}$ at 7 mm wavelength, and 21 nJy\,beam$^{-1}$ at 10 mm wavelength. With integrated fluxes of 37.1 mJy at 1.25 mm, 9.29 mJy for 3 mm, 0.806 mJy at 7 mm, and 0.258 mJy at 10 mm.}}
\label{fig:superearth_lstar10_rgap3_600g}
\end{figure*}

\begin{figure*}
\begin{center}
\includegraphics[scale=.45]{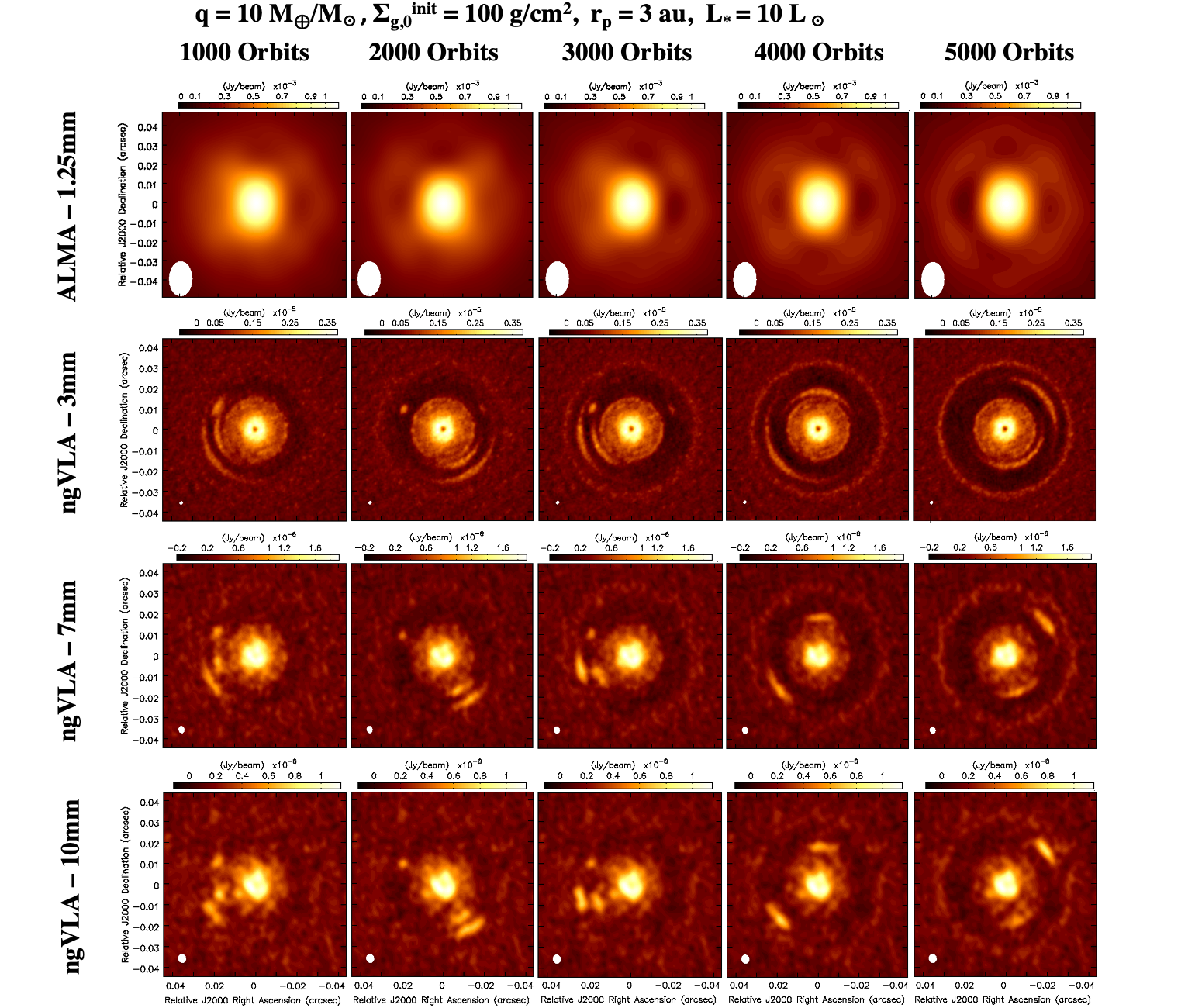}
\end{center}
\caption{\footnotesize {ALMA and ngVLA simulated observations for the dust continuum emission at 1.25 mm (ALMA), 3 mm, 7 mm, and 10 mm (ngVLA) for disk models with a planet with planet-to-star mass ratio of $10~M_{\oplus}/M_{\odot}$ at 3 au from the host star, stellar luminosity $L_{\star} = 10~L_{\odot}$, and an initial gas surface density of 100 g/cm$^2$ at the location of the planet. Each column represents the specific snapshot of the simulation in terms of number of planet orbits, from left to right of 1000, 2000, 3000, 4000, and 5000 orbits, respectively. The resulting synthesized beams have FWHM $=$ 18\,mas (2.5 au) $\times$ 12\, mas (1.7 au) at 1.25 mm,  1.7\,mas (0.23 au) $\times$ 1.3\, mas (0.19 au) at 3 mm, 3.4\,mas (0.47 au) $\times$ 2.7\, mas (0.38 au) at 7 mm, and 4.2\,mas (0.59 au) $\times$ 3.7\, mas (0.52 au) at 10 mm. The resulting rms noise levels on the maps are 2800 nJy\,beam$^{-1}$ at 1.25 mm, 86 nJy\,beam$^{-1}$ at 3 mm, 35 nJy\,beam$^{-1}$ at 7 mm, and 21 nJy\,beam$^{-1}$ at 10 mm. The disk model has integrated fluxes of 23.2 mJy at 1.25 mm, 4.07 mJy at 3 mm, 0.280 mJy at 7 mm, and 0.080 mJy at 10 mm, respectively, calculated at the 5000 orbits snapshot.}}
\label{fig:superearth_lstar10_rgap3_100g}
\end{figure*}

\begin{figure*}
\begin{center}
\includegraphics[scale=.45]{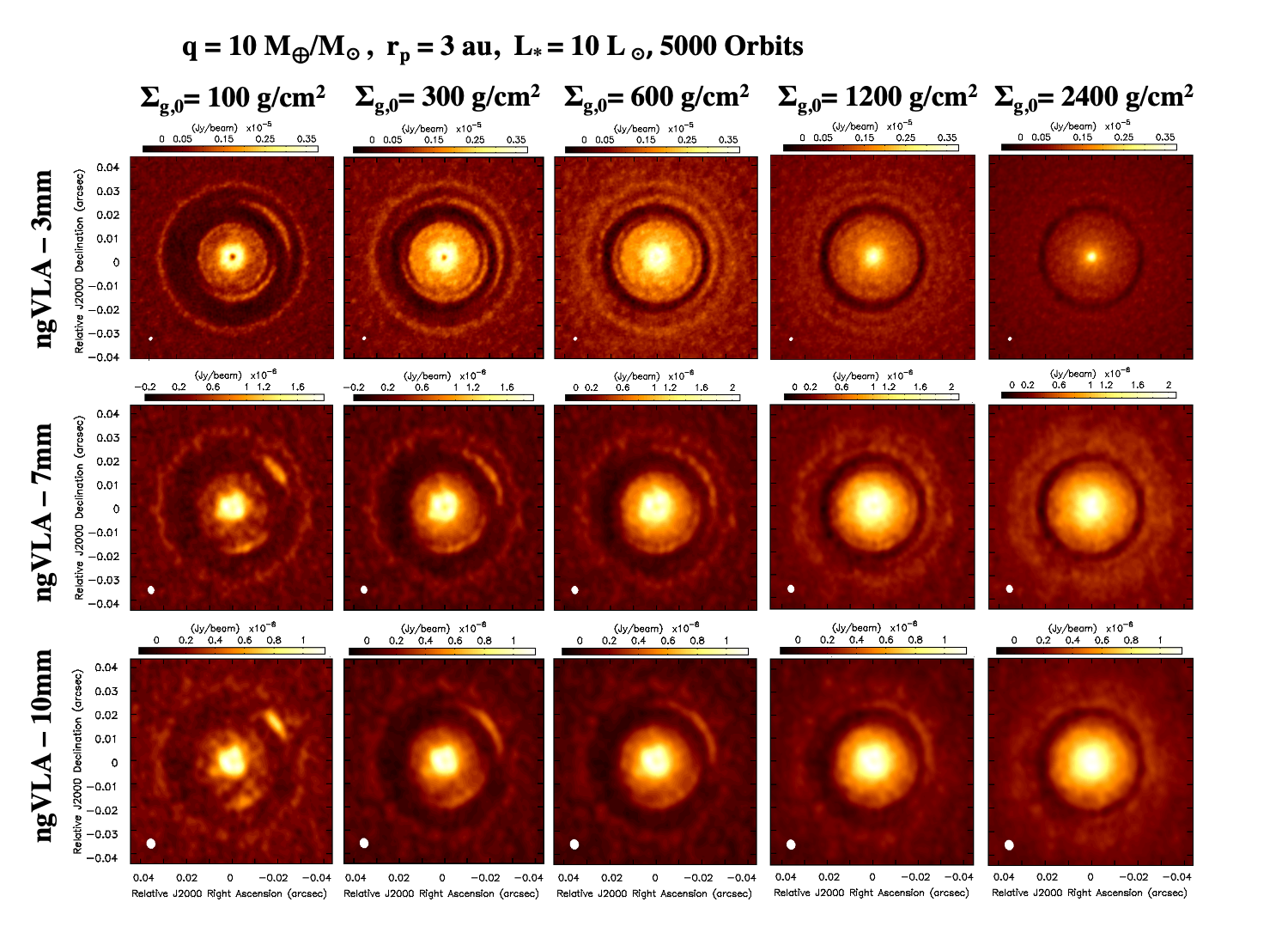}
\end{center}
\caption{\footnotesize {ALMA and ngVLA simulated observations for the dust continuum emission at 3 mm, 7 mm, and 10 mm (ngVLA) for disk models with a planet with planet-to-star mass ratio of $10~M_{\oplus}/M_{\odot}$ at 3 au from the host star, stellar luminosity $L_{\star} = 10~L_{\odot}$, at 5000 orbits. Each column shows the results for different initial gas surface densities at the planet orbital radius, as labeled on the top of each column. The resulting synthesized beams have FWHM $=$  1.7\,mas (0.23 au) $\times$ 1.3\, mas (0.19 au) at 3 mm, 3.4\,mas (0.47 au) $\times$ 2.7\, mas (0.38 au) at 7 mm, and 4.2\,mas (0.59 au) $\times$ 3.7\, mas (0.52 au) at 10 mm. The resulting rms noise levels on the maps are 2800 nJy\,beam$^{-1}$ at 1.25 mm, 86 nJy\,beam$^{-1}$ at 3 mm, 35 nJy\,beam$^{-1}$ at 7 mm, and 21 nJy\,beam$^{-1}$ at 10 mm. The disk models have integrated fluxes at 3 mm of 4.07 mJy, 6.6 mJy, 9.29 mJy, 12.2 mJy, and 14.1 mJy, at 7 mm of 0.280 mJy, 0.53 mJy, 0.806 mJy, 1.01 mJy, and 1.42 mJy, at 10 mm of 0.08 mJy, 0.16 mJy, 0.258 mJy, 0.373 mJy, and 0.521 mJy for initial surface densities between 100 g/cm$^{2}$ and 2400 g/cm$^{2}$, respectively. }}
\label{fig:superearth_lstar10_rgap3_density}
\end{figure*}

\begin{figure*}
\begin{center}
\includegraphics[scale=.6]{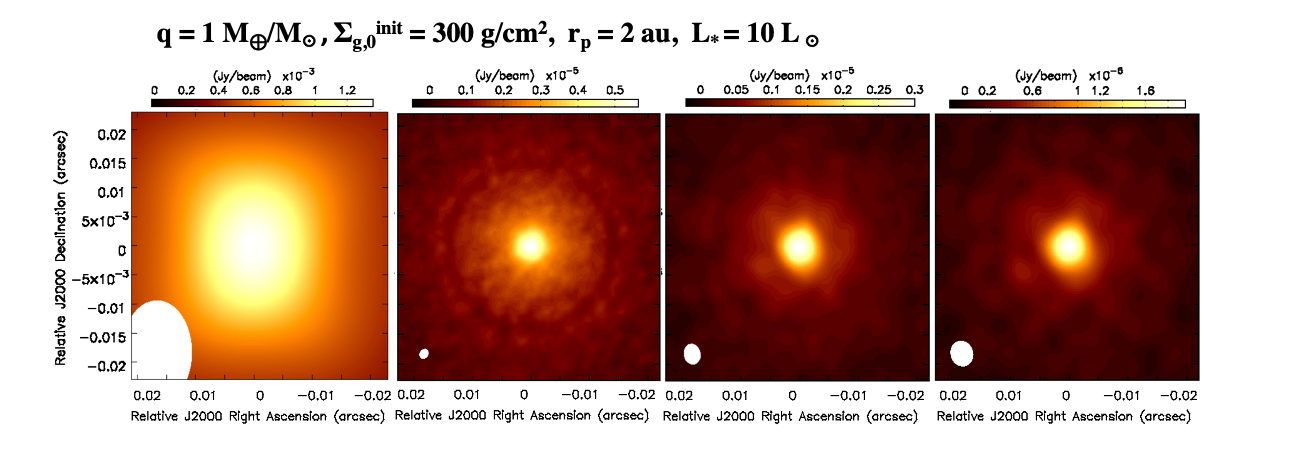}
\end{center}
\caption{\footnotesize {ALMA and ngVLA simulated observations for the dust continuum emission at 1.25 mm (ALMA), 3 mm, 7 mm, and 10 mm (ngVLA), from left to right, for a disk model with planet-to-star mass ratio of  $1~M_{\oplus}/M_{\odot}$, planet at 2 au from the host star, $L_{\star} = 10~L_{\odot}$, and an initial gas surface gas density of 300 g/cm$^2$ at the planet orbital radius. This disk model is discussed in Section~\ref{sec:results3.4}. These images were produced after 7000 planet orbits from the beginning of the simulations. The resulting synthesized beams have FWHM $=$ 18\,mas (2.5 au) $\times$ 12\, mas (1.7 au) at 1.25 mm,  1.7\,mas (0.23 au) $\times$ 1.3\, mas (0.19 au) at 3 mm, 3.4\,mas (0.47 au) $\times$ 2.7\, mas (0.38 au) at 7 mm, and 4.2\,mas (0.59 au) $\times$ 3.7\, mas (0.52 au) at 10 mm. The resulting rms noise levels on the maps are 2800 nJy\,beam$^{-1}$ at 1.25 mm, 86 nJy\,beam$^{-1}$ at 3 mm, 35 nJy\,beam$^{-1}$ at 7 mm, and 21 nJy\,beam$^{-1}$ at 10 mm. The model integrated fluxes are 26.5 mJy at 1.25 mm, 4.25 mJy at 3 mm, 0.330 mJy at 7 mm, and 0.105 mJy at 10 mm.}}
\label{fig:earthmass_rgap2}
\end{figure*}

\begin{figure*}
\begin{center}
\includegraphics[scale=.45]{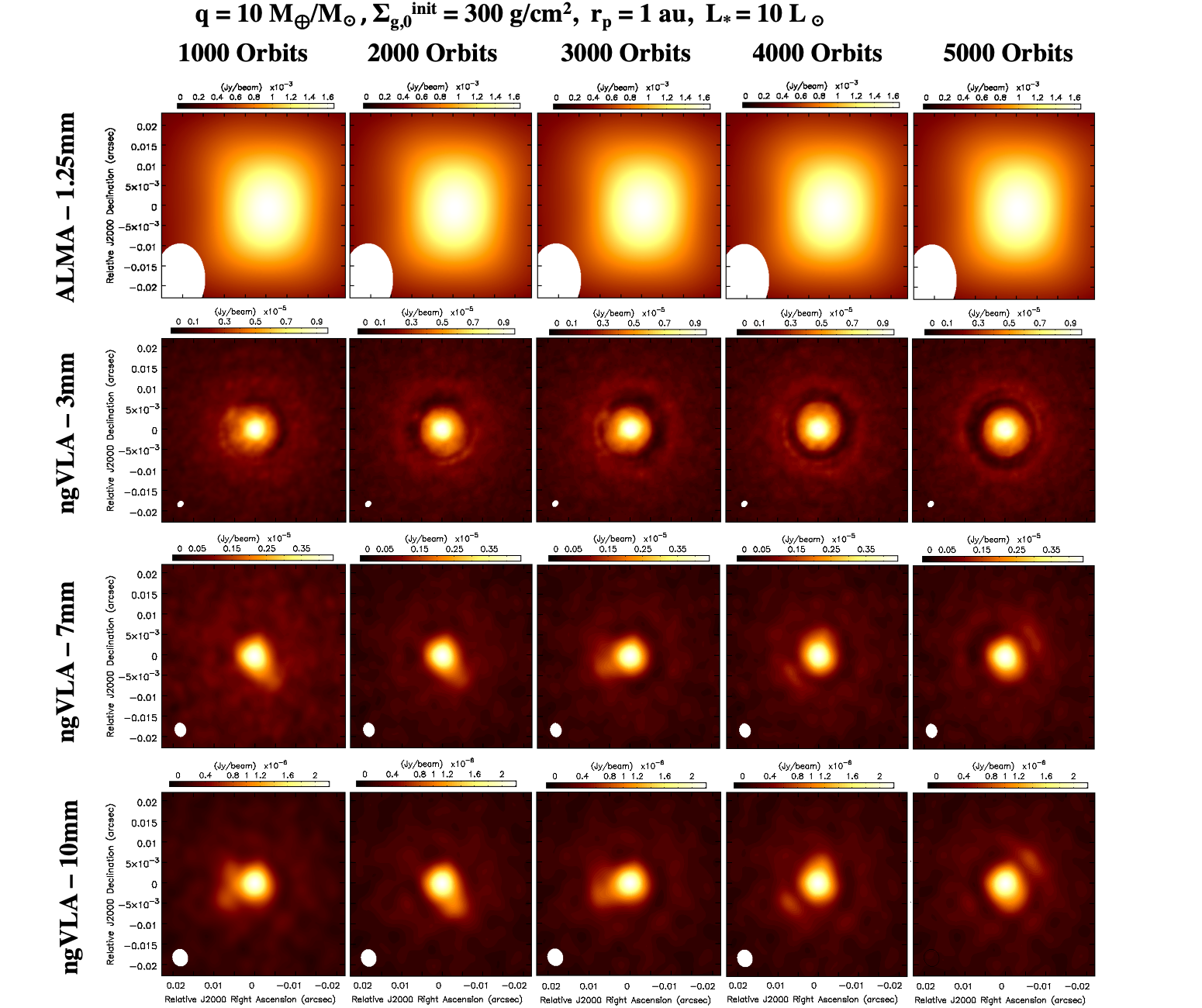}
\end{center}
\caption{\footnotesize {ALMA and ngVLA simulated observations for the dust continuum emission at 1.25 mm (ALMA), 3 mm, 7 mm, and 10 mm (ngVLA) for disk models with a planet at 1 au from the host star, planet-to-star mass ratio of $10~M_{\oplus}/M_{\odot}$, stellar luminosity $L_{\star} = 10~L_{\odot}$, and an initial gas surface density of 300 g/cm$^2$ at the planet location. Each column shows the temporal evolution with snapshots, from left to right, of 1000, 2000, 3000, 4000, and 5000 planet orbits from the beginning of the simulations, respectively. The resulting synthesized beams have FWHM $=$ 18\,mas (2.5 au) $\times$ 12\, mas (1.7 au) at 1.25 mm, 1.7\,mas (0.23 au) $\times$ 1.3\, mas (0.19 au) at 3 mm, 3.4\,mas (0.47 au) $\times$ 2.7\, mas (0.38 au) at 7 mm, and 4.2\,mas (0.59 au) $\times$ 3.7\, mas (0.52 au) at 10 mm. The resulting rms noise levels are 2800 nJy\,beam$^{-1}$ at 1.25 mm, 86 nJy\,beam$^{-1}$ at 3 mm, 35 nJy\,beam$^{-1}$ at 7 mm, and 21 nJy\,beam$^{-1}$ at 10 mm. The model integrated fluxes are 10.8 mJy at 1.25 mm, 1.62 mJy at 3  mm, 0.120 mJy at 7 mm, and 0.048 mJy at 10 mm, respectively, calculated at the 5000 orbits snapshot.}}
\label{fig:superearthmass_Lstar10_rgap1}
\end{figure*}

\begin{figure*}
\begin{center}
\includegraphics[scale=.45]{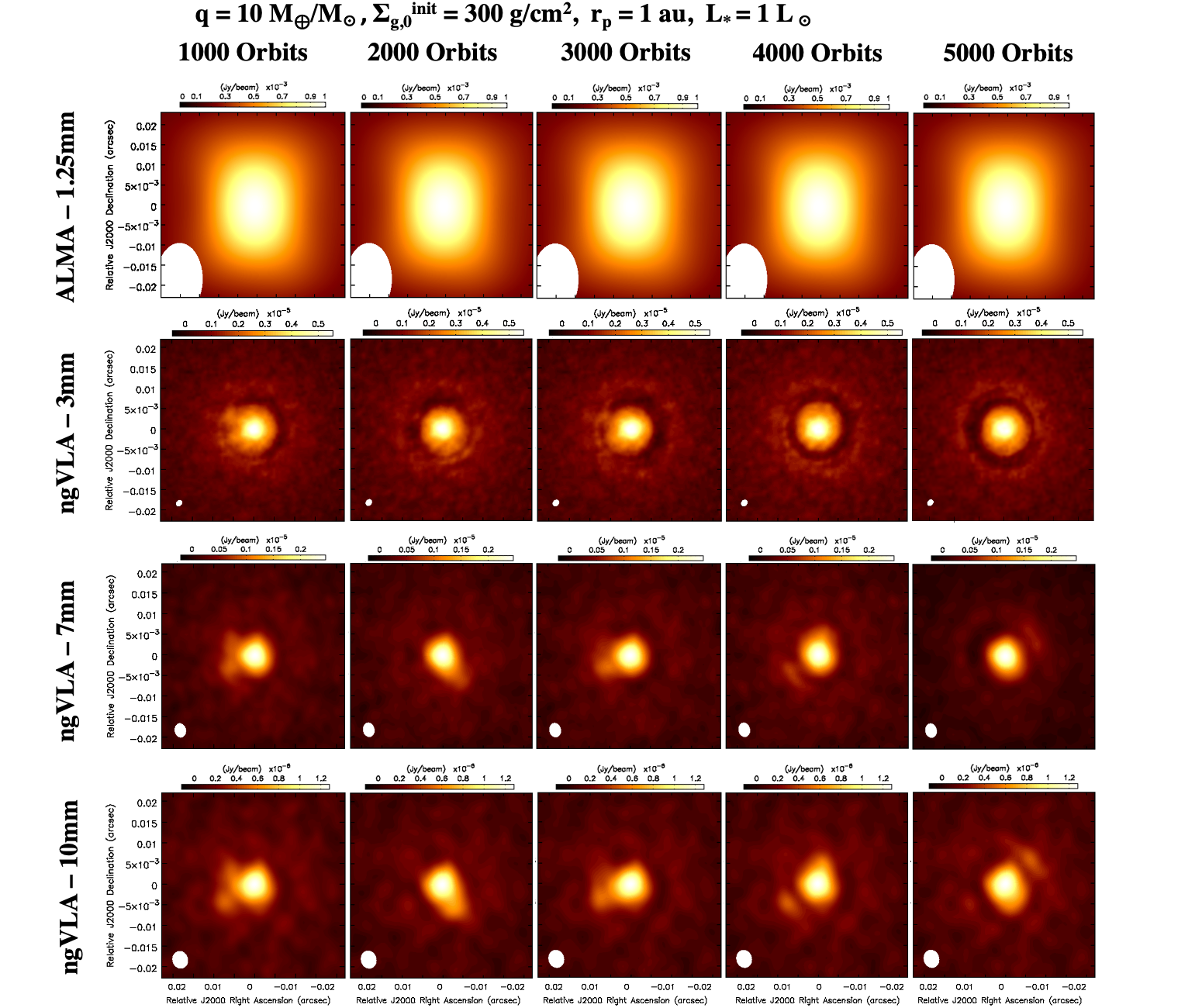}
\end{center}
\caption{\footnotesize {ALMA and ngVLA simulated observations for the dust continuum emission at 1.25 mm (ALMA), 3 mm, 7 mm, and 10 mm (ngVLA) for disk models with a planet at 1 au from the star, with planet-to-star mass ratio of $10~M_{\oplus}/M_{\odot}$, stellar luminosity $L_{\star} = 1~L_{\odot}$, and an initial gas surface density of 300 g/cm$^2$ at the planet location. Each column represents the specific snapshot of the simulation in terms of number of planet orbits, from left to right, of 1000, 2000, 3000, 4000, and 5000 orbits, respectively. The resulting synthesized beams have FWHM $=$ 18\,mas (2.5 au) $\times$ 12\, mas (1.7 au) at 1.25 mm,  1.7\,mas (0.23 au) $\times$ 1.3\, mas (0.19 au) at 3 mm, 3.4\,mas (0.47 au) $\times$ 2.7\, mas (0.38 au) at 7 mm, and 4.2\,mas (0.59 au) $\times$ 3.7\, mas (0.52 au) at 10 mm. The resulting rms noise levels on the maps are 2800 nJy\,beam$^{-1}$ at 1.25 mm, 86 nJy\,beam$^{-1}$ at 3 mm, 35 nJy\,beam$^{-1}$ at 7 mm, and 21 nJy\,beam$^{-1}$ at 10 mm. The model has integrated fluxes of 5.86 mJy at 1.25 mm, 1.02 mJy at 3 mm, 0.082 mJy at 7 mm, and 0.025 mJy at 10 mm.}}
\label{fig:superearth_Lstar1_rgap1}
\end{figure*}

\begin{figure*}
\begin{center}
\includegraphics[scale=.6]{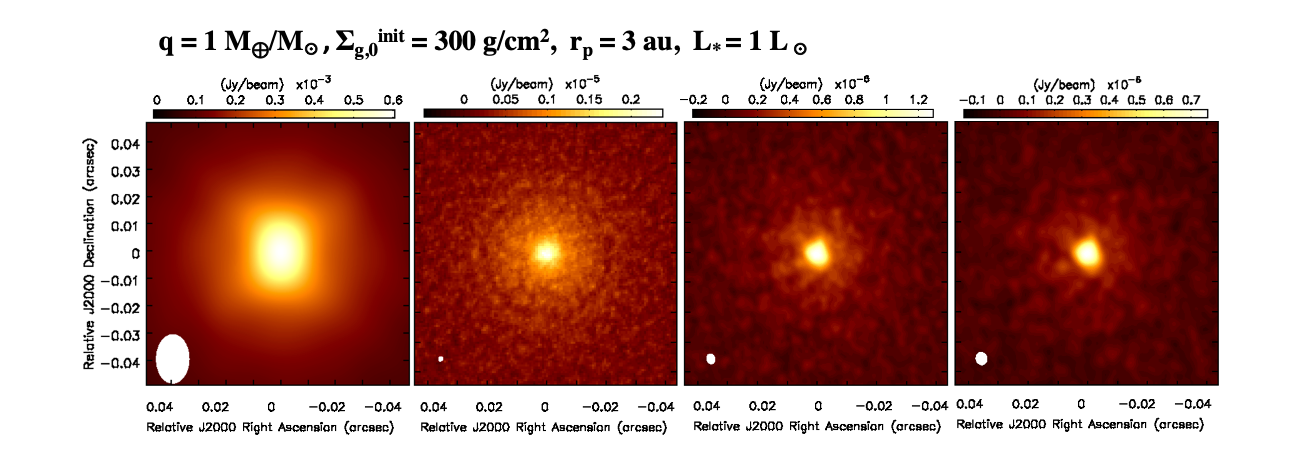}
\end{center}
\caption{\footnotesize {ALMA and ngVLA simulated observations for the dust continuum emission at 1.25 mm (ALMA), 3 mm, 7 mm, and 10 mm (ngVLA) for the disk model with a planet with planet-to-star mass ratio of $1~M_{\oplus}/M_{\odot}$ at 3 au from the host star, stellar luminosity $L_{\star} = 1~L_{\odot}$, and an initial gas surface density of 300 g/cm$^2$ at the planet location. This model is presented in Section~\ref{sec:results3.5}. Each panel is for the snapshot at 7000 planet orbits from the beginning of the simulation. The resulting synthesized beams have FWHM $=$ 18\,mas (2.5 au) $\times$ 12\, mas (1.7 au) at 1.25 mm,  1.7\,mas (0.23 au) $\times$ 1.3\, mas (0.19 au) at 3 mm, 3.4\,mas (0.47 au) $\times$ 2.7\, mas (0.38 au) at 7 mm, and 4.2\,mas (0.59 au) $\times$ 3.7\, mas (0.52 au) at 10 mm. The resulting rms noise levels on the maps are 2800 nJy\,beam$^{-1}$ at 1.25 mm, 86 nJy\,beam$^{-1}$ at 3 mm, 35 nJy\,beam$^{-1}$ at 7 mm, and 21 nJy\,beam$^{-1}$ at 10 mm. The disk model have integrated fluxes of 15.1 mJy at 1.25 mm, 4.10 mJy at 3 mm, 0.386 mJy at 7 mm, and 0.120 mJy at 10 mm.}}
\label{fig:earthmass_Lstar1_rgap3}
\end{figure*}

We start discussing the results of our simulations for one specific model, which, for simplicity, we will refer to as the \textit{reference model}. This was calculated with a planet at an orbital radius of 3 au and with a planet-to-star mass ratio of $10~M_{\oplus}/M_{\odot}$, in a disk with an initial gas surface density of 300 g/cm$^{2}$ at the orbital radius of the planet. The star in this reference model has a stellar luminosity $L_{\star} = 10~L_{\odot}$.

The left column of Figure~\ref{fig:gasdustdens} shows the time evolution of the azimuthally-averaged radial profiles of the gas and dust surface densities, and of the local dust-to-gas mass ratio for this reference model. In particular, the top panel shows how the effects of the disk-planet interaction are evident already after 1000 planet orbits, and some evolution of the gas density is still present at $4000-5000$ orbits. As already reported in several previous studies for the disk-planet interaction in low-viscosity disks \citep[e.g.,][]{zhu14,bae16,bae17,Dong:2017,Miranda:2019}, multiple local maxima and minima are found in the radial profile of the gas density outside of the location of the main gap centered at the planet orbital radius, with a magnitude of these fluctuations strongly dependent on the distance from the planet.

As a consequence of the aerodynamic coupling between gas and dust, and the related radial drift of grains due to the drag by the gas rotating at non-Keplerian speeds \citep[e.g.,][]{weidenschilling77,brauer08}, the perturbations on the densities of dust have significantly higher amplitudes than those for the gas. This results in multiple dusty rings which become more and more prominent over time. Another consequence is that the local dust-to-gas mass ratio oscillates between values well above 0.01 in the rings, down to values well below 0.01 in the gaps. For reference, the Stokes parameter of grains with sizes of 1 mm and 10 mm are of the order of $10^{-4}$ and $10^{-3}$, respectively, at the beginning of the simulation for the reference model.  

Although the radial profiles in Figure~\ref{fig:gasdustdens} are useful to visualize and quantify the effects of the planet-disk interactions on both the gas and dust components of the disk, it is important to notice that the reference model discussed in this section contains azimuthally asymmetric structures. These are particularly evident in the images of the dust continuum emission at 3, 7 and 10 mm (bottom row in Figure~\ref{fig:models}). These structures are due to dust drifting toward local pressure and density maxima produced by Rossby Wave instabilities (RWIs) at the gap edges \citep[e.g.,][]{Lovelace:1999,Li:2000,Li:2001,Lyra:2019}. The fact that these structures are better visible at longer wavelengths reflects the more efficient drift for grains of larger sizes, which dominate the emission at longer wavelengths, and make the azimuthal extent of the arc-like structure narrower for the dust emission at lower frequencies. Another important factor is the lower optical depth of the dust emission at longer wavelengths which makes the surface brightness more sensitive to the spatial variation of dust density, hence increasing the contrast on the map. For example, in the reference model after 5000 planet orbits, the optical depth of the dust emission near the peaks of the rings is $\approx 20$ at 1 mm and $\approx 1$ at 1 cm. For the model discussed in the next section with $q = 1~M_{\oplus}/M_{\odot}$ the optical depth is $\approx 7$ at 1 mm and $\approx 0.4$ at 1 cm.

These structures are visible in
Figure~\ref{fig:superearth_lstar10_rgap3_300g}, which showcases the results of the ALMA and ngVLA observations of the reference model for different snapshots of the simulations. 
Whereas the gap structure is just barely visible in the ALMA map at 1.25 mm, the ngVLA observations are able to detect and spatially resolve several radial and azimuthal structures in the disk.
 
The comparison between maps for different snapshots also highlight a clear evolutionary trend, i.e. nearly all the dusty structures tend to become more and more pronounced with time, as a consequence of the radial drift of grains discussed above.

For the first few thousand planet orbits, compact emission from dust in the Lagrangian points L4 and L5 of the star-planet system is also detected (and with best visibility at 3 mm), but already after 4000 orbits the emission gets below the sensitivity limits of the simulated observations.
These features are also visible in the radial profile of the dust surface density shown in Figure~\ref{fig:gasdustdens}. A depletion timescale of few thousand orbits for dust in these Lagrangian points can have important consequences for the Trojan dust and larger Trojan bodies expected for terrestrial planets and Super Earths \citep[see][for a discussion of Trojan bodies of giant planets, and possible implications for the Trojan solids found in the Solar System]{Montesinos:2020}.

\subsection{Models with lower planet-to-star mass ratio}
\label{sec:results3.2}

After the presentation of the reference model, the first parameter we vary is the planet-to-star mass ratio. 
The right column of Figure~\ref{fig:gasdustdens} shows the radial profiles of the gas and dust density across the disk for a disk model with the same parameters as the reference model, except for a planet-to-star mass ratio 10 times lower, i.e. $q = 1~M_{\oplus}/M_{\odot}$, while keeping the planet orbital radius at 3 au from the star. Because of the much weaker gravity of the planet, the gravitational torques on the disk are weaker as well, and the rings/gaps structures are much less pronounced than in the reference model, both in terms of the contrast between rings and gaps and radial width of the gaps. 

At the same time, as for the reference model, the radial drift experienced by grains significantly amplifies the structures in the dust component. For example, after 7000 planet orbits, the dust surface density at the location of the planet is about $1~\rm{g/cm}^2$, whereas towards the peaks in the inner and outer rings it reaches $\approx 8-10~\rm{g/cm}^2$. In the gap, the dust-to-gas mass ratio goes down to 0.003, while in the rings it reaches values of $0.02-0.03$, hence with oscillations of about $2-3\times$ below and above the initial ratio of 0.01. 

Another important difference with the reference model is that the less steep radial gradient in the gas density does not produce RWIs at the gap edges, and, as a consequence, the disk structures are much more azimuthally symmetric.
This can be seen in Figure~\ref{fig:earth_lstar10_rgap3_300g}, which showcases the results of the ALMA and ngVLA observations for the disk model with $q = 1~M_{\oplus}/M_{\odot}$.
For this model, the gap opened by the planet is detected by the ngVLA, with decreasing signal-to-noise ratios at longer wavelengths, similarly to the reference model, whereas ALMA fails to detect the gap as a consequence of the much larger synthesized beam. 
In the case of the ngVLA map at 3 mm and at 7000 orbits, the peak signal-to-noise ratio on the inward and outward edges of the gap itself are 12.1 and 11.2, respectively.

We also notice that the ring/gap structure is not visible within $1000-2000$ planet orbits from the beginning of the simulations. Only after 2000 planet orbits the gap starts to be marginally detected with the ngVLA at 3 mm, and the detection becomes more and more significant after $3000-4000$ orbits. 

\subsection{Models with different gas surface densities}
\label{sec:results3.3}

We present here the results of the models in which we varied the initial gas surface density of the disk. With a fixed initial dust-to-gas mass ratio of 100, varying the gas density automatically modifies also the dust density. Moreover, varying the gas density also changes the aerodynamic coupling between the gas and dust grains (see above the definition of the Stokes parameter), with an influence on the timescales of radial drift of solids across the disk.
The reference model presented in Section~\ref{sec:results3.1} considered an initial gas surface density of 300 g/cm$^2$ at the planet location ($r_p =$ 3 au). Here we discuss the results of simulations with values of 100, 600, 1200, and 2400 g/cm$^2$ for this parameter. 

Figure~\ref{fig:depletion} displays the $\Delta$ parameter, which quantifies the amount of dust within the gap relative to the amount of dust in the rings, as a function of time. The plots show the results of simulations with different values of the initial gas surface density, and for values of the planet-to-star mass ratio of $1~M_{\oplus}/M_{\odot}$ (left plot) and $10~M_{\oplus}/M_{\odot}$ (right). More precisely, the $\Delta$ parameter is defined as the ratio between the average dust density in the gap at the orbital radius of the planet and the average density of the peaks of the two closest rings from the planet, one inside and one outside of the planet orbit. With this definition, $\Delta$ depends both on the dust density in the gap and in the inner and outer rings, but, as shown in Fig.~\ref{fig:gasdustdens}, the variation of the density of dust corotating with the planet is much larger than the density in the rings (especially for the case with $10~M_{\oplus}/M_{\odot}$).

The comparison between the two panels in Figure~\ref{fig:depletion} highlights the very different timescales in the depletion of the gap in systems with different planet-to-star mass ratios. 
In the simulations with $q=10~M_{\oplus}/M_{\odot}$, the $\Delta$ parameter reaches values $\sim 0.2 - 0.7$ within the first $1000-3000$ planet orbits of evolution, and then strongly decreases by factors of $\sim 10$ and $\sim 100$ at 4000 and 5000 orbits, respectively. Conversely, the temporal evolution is much more gradual in the case of the simulations with  $q=1~M_{\oplus}/M_{\odot}$. Here the weaker gravity of the planet takes much longer to deplete the gap, as even after 7000 planet $\Delta \sim 0.1 - 0.2$, and the decrease with time is roughly linear. In terms of the dependence on the gas surface density of the disk, although the qualitative trend does not depend on density, one can notice faster evolution for the disk models with lower gas densities. This is in line with the expectation of a faster radial drift of grains with larger (but lower than 1) values of the Stokes parameter, since the Stokes parameter is inversely proportional to the local density of gas in the disk. 

The simulated observations for the disk models with initial gas surface densities at 3 au of 600 and 100 g/cm$^2$ are shown in Fig.~\ref{fig:superearth_lstar10_rgap3_600g} and \ref{fig:superearth_lstar10_rgap3_100g}, respectively. The main difference between these results, including also the results for the reference model with gas density of 300 g/cm$^2$ shown in Figure~\ref{fig:superearth_lstar10_rgap3_300g}, is that the models with lower gas densities have generally more pronounced substructures: gaps are radially wider and with higher depletion, and the azimuthally asymmetric structures related to local dust trapping are also more concentrated along the azimuthal direction, including also in the Lagrangian points L4 and L5 \citep{Lyra:2009}. Similarly to the discussion above for the depletion in the planet-induced gaps, this is an effect of the faster drift of grains with higher values of the Stokes parameter. Despite the fact that dust substructures tend to be more prominent in disk models with lower gas densities, the correspondingly lower densities in dust make some of these structures detectable only at marginal signal-to-noise ratios, especially at 10 mm (Fig.~\ref{fig:superearth_lstar10_rgap3_100g}). At the same time, the higher optical depths for the models with the highest densities considered in this work limit the visibility of the azimuthal structures even at 10 mm.

\subsection{Models with different planet orbital radii}
\label{sec:results3.4}

Another model parameter that was varied in our simulations is the orbital radius of the planet. This parameter has an effect on the substructures generated from the disk-planet interactions. For example, the radial width of the main gap increases linearly with the orbital radius of the planet \citep[e.g.,][]{Ricci:2018}. 

Whereas the reference model presented in Section~\ref{sec:results3.1} considered a planet at 3 au from the star, we also ran simulations for a disk model with $q = 1~M_{\oplus}/M_{\odot}$ with the planet at 2 au, and for a disk model with $q = 10~M_{\oplus}/M_{\odot}$ with the planet at 1 au.
 
Figure~\ref{fig:earthmass_rgap2} displays the results of the ALMA and ngVLA simulations for models with $q = 1~M_{\oplus}/M_{\odot}$ and planet at 2 au. The maps show that the gap is only detectable at wavelengths of 3 and 7 mm, barely detectable at 10 mm, and not detected with ALMA. The same gap is only marginally detected even at 3 mm in the snapshots with planet orbits lower than 7000. We also ran simulations for a planet with the same $q-$ratio at 1 au from the star, but the gap was not detectable with neither ngVLA nor ALMA, and for that reason those results are not shown here.

For the model with planet at 1 au and $q = 10~M_{\oplus}/M_{\odot}$, Figure~\ref{fig:superearthmass_Lstar10_rgap1} showcases the temporal evolution of the dust emission at wavelengths of 1.25 mm, 3 mm, 7 mm, and 10 mm, respectively. As for the previous case discussed in this section, the ALMA observations lack the angular resolution to detect the gap opened by the planet. The situation is different for the ngVLA, which can instead detect both the main gap structure as well as azimuthally asimmetric structures, especially at 3 mm.
The gap structure and the arc-like structure is detected, although at lower signal-to-noise ratios, also at 7 and 10 mm with the ngVLA, with the visibility of these structures significantly increasing after few thousand planet orbits of evolution.

\subsection{Models with lower stellar luminosity}
\label{sec:results3.5}

In all the simulations presented so far, the bolometric luminosity of the star was assumed to be $L_{\star} = 10~L_{\odot}$. For several pre-Main Sequence evolutionary models, a young star with the mass of the Sun reaches this luminosity within its first Myr of age \citep[e.g.,][]{Siess:2000}. In this section we present the results of simulations 
with a lower stellar luminosity of $L_{\star} = 1~L_{\odot}$. In this work, this change affects only the dust temperature which is used for the computation of the model images for the dust continuum emission, but not the gas temperature, and pressure, used in the hydrodynamical simulations. We also note that, at a fixed stellar mass, the dependence of the vertical scale height with stellar luminosity is rather weak. For example, a disk around a Solar-mass star with $L_{\star} = 1~L_{\odot}$ has an aspect ratio $h/r = 0.03$ at 3 au assuming $\phi = 0.02$ (Eq.~\ref{eq:temp}). If $L_{\star} = 10~L_{\odot}$, then $h/r = 0.04$.   Hence, the model images with lower stellar luminosity are characterized by a lower surface brightness as a consequence of the lower dust temperature.  

Figure~\ref{fig:superearth_Lstar1_rgap1} displays the results of the model with $q = 10~M_{\oplus}/M_{\odot}$ and with planet at 1 au from a star with $L_{\star} = 1~L_{\odot}$.  These simulations are similar to the case with $L_{\star} = 10~L_{\odot}$ shown in Fig.~\ref{fig:superearthmass_Lstar10_rgap1}, but have lower surface brightnesses by a factor of $\approx 2$. Even with these lower fluxes, the gap is still detectable with the ngVLA, with the usual trends of higher signal-to-noise ratios at 3 mm and towards the later snapshots of the simulations. 

Figure~\ref{fig:earthmass_Lstar1_rgap3} shows the results of the simulations after 7000 planet orbits for a planet at 3 au from the star with $q = 1~M_{\oplus}/M_{\odot}$, and again with stellar luminosity $L_{\star} = 1~L_{\odot}$. The planet-induced gap is noticeable with the ngVLA at 3 mm and at 7 mm, even though with marginal signal-to-noise ratio.

\section{Discussion}
\label{sec:discussion}


\begin{figure*}
\begin{center}
\includegraphics[scale=.6]{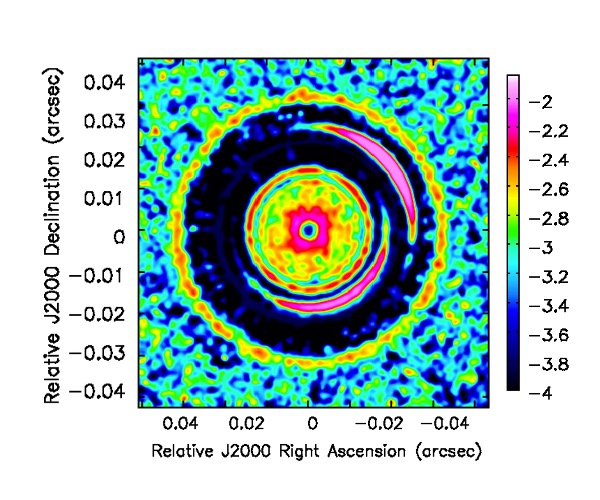}
\end{center}
\caption{\footnotesize {Map for the spectral index of the dust emission between 3 and 7 mm for the reference model presented in Section~\ref{sec:results3.1}. This was derived by combining the simulated ngVLA images at 3 and 7 mm after 5000 planet orbits, shown in Figure~\ref{fig:superearth_lstar10_rgap3_300g}.}}
\label{fig:spectralindex}
\end{figure*}

In Section~\ref{sec:results} we have shown that the ngVLA can resolve several substructures due to the disk-planet interaction in low viscosity disk systems in nearby star forming regions, with $q = 1 - 10~M_{\oplus}/M_{\odot}$ and with the planet at orbital radii down to $1-2$ au from the central star. These substructures include multiple rings/gaps, arc-like structures and dust concentrated in Lagrangian points of the planet-star system. Although the signal-to-noise ratios decrease at longer wavelengths as a consequence of the coarser angular resolution and weaker dust emission at lower frequencies, our results show that for several models, especially those with $q = 10~M_{\oplus}/M_{\odot}$, the combined observations at multiple wavelengths, between 3 mm and 1 cm in this study, can highlight variations in the morphology of some of these structures across wavelength. This is a result of the spatial segregation of grains with different sizes, as expected by the mechanism of radial drift of solids in a gaseous disk~\citep[e.g.,][]{Laibe:2012,Laibe:2014a,Laibe:2014b,Birnstiel:2016}.

The variation across wavelength of the disk emission can be diagnosed also through the map of the spectral index $\alpha_{\rm{mm}}$ of the dust emission at mm wavelengths ($F_{\nu} \propto \lambda^{\alpha_{\rm{mm}}}$). Because of the expected anti-correlation between the absolute value of $\alpha_{\rm{mm}}$ and grain size, at least for grain sizes larger than $\sim 0.1 - 1$ mm \citep[e.g.,][]{Ricci:2010}, one would expect to find the lowest absolute values on the spectral index map in correspondence to the dust substructures formed via the trapping of the larger grains, as discussed above \citep{vandermarel:2015,Barge:2017}. This is confirmed in Figure~\ref{fig:spectralindex}, which shows absolute values of $\alpha_{\rm{mm}}$ as low as $\approx 1.8 - 2.2$ at the locations of the dust substructures found in the reference model presented in Section~\ref{sec:results3.1}.

It is important to emphasize that for several of the models presented here, especially the ones with the lower $q$-value of $1~M_{\oplus}/M_{\odot}$, these substructures start to become detectable in the dust continuum maps only after several thousand planet orbits. This demonstrates the importance of simulations that span at least several thousand orbital timescales of evolution, especially when investigating the disk-planet interaction with low-mass planets like Earth-like or Super Earth planets. This also explains the apparent discrepancy between the results presented here and those from previous similar investigations, such as for example in \citet{Ricci:2018}, in which hydrodynamical simulations were run only for less than 2000 planet orbits.
This result is particularly interesting also considering that Super-Earth and Earth-mass planets are more commonly found around Main Sequence stars than higher mass giant planets \citep[e.g.,][]{Howard:2010, Petigura:2013}.

The simulations of the observations presented in this work assumed a distance of 140 pc for the disks. This is similar to the average distance of young stars in nearby regions such as Taurus~\citep[e.g.,][]{Kenyon:2008} or Ophiuchus~\citep[e.g.,][]{Wilking:2008}.
\citet{Andrews:2013} performed a survey in Taurus of 227 young stars surrounded by disks using the Submillimeter Array (SMA) at a wavelength of 1.3 mm, which is comparable to the wavelength of 1.25 mm for our ALMA simulations. 80 disks of those surveyed, have a flux density at 1.3 mm above 10 mJy, which is close to the flux density value of the faintest models in our investigation. An ALMA survey of 147 disks was presented by \citet{Cieza:2019} in the Ophiuchus star forming region as part of the \textit{Ophiuchus Disc Survey Employing ALMA} (ODISEA) program. They found 60 disks with integrated fluxes greater than 10 mJy. 

 
Another potentially interesting investigation that the ngVLA observations would allow is the study of the proper motion of the azimuthally asymmetric structures, which are nearly always detected in our simulations with $q = 10~M_{\oplus}/M_{\odot}$. For example, given the Keplerian orbital velocity at 1 au from a Solar-mass star, and a spatial resolution of 0.2 au from the ngVLA at 3 mm for a disk at 140 pc, the proper motion of a source detected at a signal-to-noise ratio of $\approx 5$, would be detectable at 3$\sigma$ with two epochs separated by just $3-4$ days. For a source orbiting at 3 au from a Solar-mass star the proper motion would be detectable at 3$\sigma$ with epochs separated by about 6 days.
 
We also note that our simulations lack the numerical resolution to investigate the possible presence and emission of circumplanetary disks accreting onto the planets \citep[e.g.,][]{Isella:2019}. The potential for detection of these systems by ALMA and the ngVLA has been discussed in~\citet{Zhu:2018}.  
 
Finally, although in this work we have investigated the dependence of the results on a variety of different model parameters, there are other parameters that we kept fixed for our investigation. 
For example, we fixed the $\alpha$ viscosity parameter to a value of $10^{-5}$. As shown in previous works~\citep[e.g.,][]{Fung:2014, Kanagawa:2015,Dong:2017,Ricci:2018}, gas with higher viscosity by factors $> 10-100$ than considered in this paper would make the density perturbations created by the interaction with the planet to be more smoothed out, which would have an impact on the observed structures in the dust emission. For the specific case of the disk models considered here, \citet{Zhang:2018} showed that in models with higher values for the gas viscosity $\alpha = 10^{-3}$ and $10^{-4}$, gaps are opened in the case of $q \approx 10~M_{\oplus}/M_{\odot}$ (see their Fig. 6 and 7), but not for $q \approx 1~M_{\oplus}/M_{\odot}$.

The simulations use a locally-isothermal equation of state. Recent works by \citet{Miranda:2019,Miranda:2020} and \citet{Zhang:2020} find that cooling can affect the shape of the gaps. In the inner disk regions considered in this study, the cooling timescale can be longer than the orbital timescale. Under these conditions, considering an adiabatic equation of state can have significant effects on the results of this work, and this will be the subject of a follow-up work in the future. 

Another parameter that we kept fixed in our simulations is the vertical aspect ratio of the disk: $h/r~(r_{p}) = 0.03$. The vertical scale height sets the strength of the pressure forces in the gas component of the disk, which, similarly to viscosity, act \textit{against} the formation of density perturbations in the disk.
We also did not investigate the dependence of our results on the grain size distribution of dust in the disk. In our simulations we assumed an initial grain size distribution $n(s) \propto s^{-3.5}$ with a maximum grain size of 1 cm~\citep[see][]{Zhang:2018}. We postpone the study on the impact of these assumptions to a future work.

\section{Conclusions}
\label{sec:conclusions}

In this work we have investigated and tested the imaging capabilities of the ngVLA, and the comparison with ALMA, for resolving gaps and rings and other substructures in the dust continuum of disks with low viscosity due to low-mass terrestrial planets. We produced models with $q = 1~M_{\oplus}/M_{\odot}$, corresponding for example to an Earth-mass planet around a Solar-mass star or a Mars-mass planet around a $0.1~M_{\odot}$ M-type star, and with $q = 10~M_{\oplus}/M_{\odot}$, corresponding to a Super-Earth around a Solar-mass star or an Earth-mass planet around a 0.1 $M_{\odot}$ M-type star \footnote{At the same time, it is important to remind the reader that our simulations were run with a fixed radial profile of the vertical scale height which is more consistent with the observed properties of disks around young stars with masses $\approx 0.5 - 1.0~M_{\odot}$.}.  

We showed that after several thousand orbits of evolution in these systems, terrestrial planets with orbital radii of $1 - 3$ au can leave imprints in the dust continuum emission that the ngVLA would be able to detect. Deep observations, especially at 3 mm, would allow us to well characterize the radial structure of the main rings/gaps which are found in several disk models with $q = 1$ and $10~M_{\oplus}/M_{\odot}$, and also detect emission from azimuthal structures close to the planet for models with $q = 10~M_{\oplus}/M_{\odot}$ which are predicted by the physics of the planet-disk interaction in disks with low viscosity.

Furthermore, multi-wavelength observations within the range of frequencies covered by the ngVLA would allow us to map the emission dominated by grains with different sizes within regions of local dust concentration. This would provide further observational constraints to the spatial segregation of grains with different sizes, to the measurement of the dust optical depth at these wavelengths, and finally to further constrain the physical process of planet-disk interaction and of grain growth in high-density regions which could be suitable for the formation of planetesimals.   

Given the very high angular resolution and astrometric precision expected for the ngVLA, together with the short orbital timescales at few au from young stars, the proper motion due to the orbital motion expected for these structure should be detectable already with observations separated by few days. This would be important also to confirm the physical association of compact structures to the disk, rather than to background (or foreground) sources which may be expected at the very high sensitivities probed by the ngVLA.

Our results show that a ngVLA with the design considered in recent studies has the potential to transform the fields of planet formation and protoplanetary disks, allowing for imaging of the disk physical structures due to low-mass terrestrial planets at sub-au resolution in nearby star forming regions. 

Although this work presents results over a relatively broad range of the parameter space of our models (planet-to-star mass ratio, planet orbital radius, disk density, stellar luminosity), there are other assumptions that can have an important impact on the results discussed here (e.g., gas viscosity, grain size distribution, thermodynamics). Further investigations over a broader region of the model parameter space will be provided in a future work.

\vskip 0.1in

\acknowledgements We thank the anonymous referee for his/her helpful comments. This work was
supported in part by the ngVLA Community Studies program, coordinated by the National Radio Astronomy Observatory,
which is a facility of the National Science Foundation operated
under cooperative agreement by Associated Universities, Inc.
\bibliographystyle{aasjournal}
\begin{singlespace}
\bibliography{main}
\end{singlespace}


\end{document}